\documentclass[galaxies,article,accept,pdftex,moreauthors]{Definitions/mdpi}

\firstpage{1} 
\pubvolume{1}
\issuenum{1}
\articlenumber{0}
\pubyear{2022}
\copyrightyear{2022}
\datereceived{6 November 2022} 
\dateaccepted{25 November 2022} 
\datepublished{} 
\hreflink{https://doi.org/}

\Title{Expectations for Horizon-Scale Supermassive Black Hole Population Studies with the ngEHT}

\TitleCitation{Expectations for Horizon-Scale Supermassive Black Hole Population Studies with the ngEHT}

\Author{Dominic~W.~Pesce $^{1,2,*}$\orcidA{}, Daniel~C.~M.~Palumbo $^{1,2}$\orcidB{}, Angelo~Ricarte $^{1,2}$\orcidC{}, Avery~E.~Broderick $^{3,4,5}$\orcidH{}, Michael~D.~Johnson $^{1,2}$\orcidD{}, Neil~M.~Nagar $^{6}$\orcidE{}, Priyamvada~Natarajan $^{2,7,8}$\orcidG{} and Jos\'e~L.~G\'omez $^{9}$\orcidF{}}

\AuthorNames{Dominic~W.~Pesce, Daniel~C.~M.~Palumbo, Angelo~Ricarte, Avery~E.~Broderick, Michael~D.~Johnson, Neil~M.~Nagar, Jos\'e~L.~G\'omez, Priyamvada~Natarajan}

\AuthorCitation{Pesce, D.; Palumbo, D.C.M.; Ricarte, A.; Broderick, A.E.; Johnson, M.D.; Nagar, N.M.; Natarajan, P.; G\'omez, J.L.}
\address{%
$^{1}$ \quad Center for Astrophysics $|$ Harvard \& Smithsonian, 60 Garden Street, Cambridge, MA 02138, USA; daniel.palumbo@cfa.harvard.edu (D.C.M.P.); angelo.ricarte@cfa.harvard.edu (A.R.); mjohnson@cfa.harvard.edu (M.D.J.) \\
$^{2}$ \quad Black Hole Initiative at Harvard University, 20 Garden Street, Cambridge, MA 02138, USA; priyamvada.natarajan@yale.edu\\
$^{3}$ \quad Perimeter Institute for Theoretical Physics, 31 Caroline Street North, Waterloo, ON N2L 2Y5, Canada; abroderick@perimeterinstitute.ca\\
$^{4}$ \quad Department of Physics and Astronomy, University of Waterloo, 200 University Avenue West, \mbox{Waterloo, ON N2L 3G1, Canada}\\
$^{5}$ \quad Waterloo Centre for Astrophysics, University of Waterloo, Waterloo, ON N2L 3G1, Canada\\
$^{6}$ \quad Astronomy Department, Universidad de Concepci\'on, Casilla 160-C, Concepci\'on, Chile; nagar@astro-udec.cl\\
$^{7}$ \quad Department of Astronomy, Yale University, 52 Hillhouse Avenue, New Haven, CT 06511, USA\\
$^{8}$ \quad Department of Physics, Yale University, P.O. Box 208121, New Haven, CT 06520, USA\\
$^{9}$ \quad Instituto de Astrof\'isica de Andaluc\'ia-C\'iSIC, Glorieta de la Astronom\'ia s/n, E-18008 Granada, Spain; jlgomez@iaa.es \\
}

\corres{Correspondence: dpesce@cfa.harvard.edu}

\abstract{We present estimates for the number of supermassive black holes (SMBHs) for which the next-generation Event Horizon Telescope (ngEHT) can identify the black hole ``shadow,'' along with estimates for how many black hole masses and spins the ngEHT can expect to constrain using measurements of horizon-resolved emission structure.  Building on prior theoretical studies of SMBH accretion flows and analyses carried out by the Event Horizon Telescope (EHT) collaboration, we construct a simple geometric model for the polarized emission structure around a black hole, and we associate parameters of this model with the three physical quantities of interest.  We generate a large number of realistic synthetic ngEHT datasets across different assumed source sizes and flux densities, and we estimate the precision with which our defined proxies for physical parameters could be measured from these datasets.  Under April weather conditions and using an observing frequency of 230\,GHz, we predict that a ``Phase 1'' ngEHT can potentially measure $\sim$50 black hole masses, $\sim$30 black hole spins, and $\sim$7 black hole shadows across the entire sky.}


\begin{document}

\nocite{M87PaperI}
\nocite{M87PaperII}
\nocite{M87PaperIII}
\nocite{M87PaperIV}
\nocite{M87PaperV}
\nocite{M87PaperVI}
\nocite{M87PaperVII}
\nocite{M87PaperVIII}

\nocite{SgrAPaperI}
\nocite{SgrAPaperII}
\nocite{SgrAPaperIII}
\nocite{SgrAPaperIV}
\nocite{SgrAPaperV}
\nocite{SgrAPaperVI}

\section{Introduction}\label{sec:Introduction}

The Event Horizon Telescope (EHT) observations of the supermassive black holes (SMBHs) in M87 \citep{M87PaperI,M87PaperII,M87PaperIII,M87PaperIV,M87PaperV,M87PaperVI,M87PaperVII,M87PaperVIII} and Sgr A* \citep{SgrAPaperI,SgrAPaperII,SgrAPaperIII,SgrAPaperIV,SgrAPaperV,SgrAPaperVI} are the first in a new era of horizon-scale studies of black holes.  The~primary observational signature on horizon scales is the black hole ``shadow,'' a ring-like emission structure surrounding a darker central region \citep{Falcke_2000,Narayan_2019}.  Simulations of accretion flows around SMBHs generically produce images that exhibit such shadows \citep{M87PaperV,SgrAPaperV}, which typically have a size comparable to that of the theoretical curve bounding the locus of impact parameters for photon trajectories that escape the black hole (i.e., the~``apparent shape'' of the black hole, from~\citet{Bardeen_1973}).  A~driving motivation for the EHT to pursue observations of M87* and Sgr A* was because these sources were anticipated to have the largest shadow sizes of all black holes on the sky \citep{M87PaperI}.

The next-generation Event Horizon Telescope (ngEHT) will build on the capabilities of the EHT by improving $(u,v)$-coverage through the addition of more stations to the array, increasing baseline sensitivities by using wider observing bandwidths, and~accessing finer angular resolution by observing at higher frequencies \citep{Doeleman_2019}.  A~natural question to ask is whether these improved capabilities will yield access to a larger pool of shadow-resolved SMBHs.  The~horizon-scale emission structure around a black hole encodes spacetime properties such as its mass and spin, and~the detection of a shadow is a distinct and relatively unambiguous identifier of the observed object's black hole nature.  Access to a population of shadow-resolved SMBHs would thus provide an opportunity to make uniquely direct and self-consistent measurements of such spacetime properties, with~attendant implications for studies of SMBH formation, growth, and~co-evolution with host~galaxies.

The suitability of any particular SMBH for shadow-resolving ngEHT observations depends primarily on three properties \citep{Pesce_2021}:
\begin{enumerate}
    \item the angular size of the SMBH shadow ($\theta$);
    \item the total horizon-scale flux density emitted by the source ($S_{\nu}$); and 
    \item the optical depth of the emitting material.
\end{enumerate}
\noindent 

The first of the above properties is set primarily by the mass of and distance to the black hole, while the latter two are more complex and depend also on the mass accretion rate and other physical conditions in the accretion flow.  However, the~detectability of horizon-scale structure from a SMBH does not guarantee the measurability of any particular quantity of interest; additional conditions must be met to ensure that, e.g.,~a black hole mass can be measured, or~that the ring-like structure associated with the black hole shadow can be distinguished from other possible emission~morphologies.

In this paper, we provide estimates for the number of SMBHs for which the ngEHT could plausibly make mass, spin, and~shadow measurements.  In Section~\ref{sec:MeasurableProxies}, we define observational proxies for each of these quantities of interest that can be accessed from the horizon-scale emission structure.  Section~\ref{sec:SyntheticData} describes our synthetic data generation procedure and our approach to estimating parameter measurement precision from ngEHT data. Our conditions for the measurability of each proxy are defined in Section~\ref{sec:NumberOfMeasurables}, where we also report the number of objects expected to satisfy these conditions for each quantity of interest.  We summarize and conclude in Section~\ref{sec:Summary}.  Throughout this paper, we use the results from \citet{Pesce_2021} as our baseline for how many SMBHs satisfy the above three detection criteria as a function of $\theta$ and $S_{\nu}$.
\footnote{The procedure \citet{Pesce_2021} used to determine the number of observable SMBHs involves integrating the supermassive black hole mass function (BHMF) to determine how many objects have shadow diameters larger than $\theta$, while also using a semi-analytic spectral energy distribution model and adopting an empirically motivated prescription for the SMBH Eddington ratio distribution function to restrict the objects under consideration to those that have flux densities greater than $S_{\nu}$ and accretion flows that are optically thin.  The~distribution of sources used in this paper assumes an observing frequency of 230\,GHz and a BHMF determined using the stellar mass function from \citet{Behroozi_2019} scaled according to the relation determined by \citet{Kormendy_2013} (i.e., the~``upper BHMF'' from \citet{Pesce_2021}).}

\section{Measurable Proxies for Quantities of~Interest}\label{sec:MeasurableProxies}

For a given SMBH, the~two primary quantities of scientific interest are its mass and spin, neither of which is directly observable by the ngEHT.  Instead, analyses of ngEHT observations will need to identify and measure features of the emission structure that serve as proxies for the desired quantities, or~else they will need to carry out some form of physical modeling to infer the SMBH mass and/or spin from the ngEHT data.  For~the proof-of-concept analyses presented in this paper, we pursue the former~strategy.

\subsection{Proxy for SMBH~Shadows}\label{sec:ShadowProxy}

One of the most generic predictions from simulated images of SMBHs is that the observed emission structure on event horizon scales should exhibit a ring-like morphology associated with the black hole shadow (e.g., \citep{M87PaperV,SgrAPaperV}).  Though it is possible for other processes to give rise to ring-like emission structures---e.g.,~the Einstein ring from a bright, compact emitter passing behind the black hole---in such cases the ring-like structure is expected to be transient.  For the purposes of this paper, we thus consider the observation of a ring-like emission morphology to be a proxy for verifying the object's black hole nature.  If~we can determine from ngEHT observations that the emission structure from a particular object is ring-like---i.e.,~if we can discern the shadow---then we can identify that object as a black~hole.

\subsection{Proxy for SMBH~Masses}\label{sec:MassProxy}

The mass of a SMBH sets the physical scale for its associated spacetime metric, and~all spacetime-sensitive length scales in the system should thus exhibit a proportionality with the gravitational radius,
\begin{equation}
\theta_g = \frac{G M}{c^2 D} ,
\end{equation}
\noindent with $M$ the black hole mass and $D$ its distance from Earth.  The~most observationally accessible length scale is the overall size of the emission region, which for a ring-like emission structure corresponds to the ring diameter, $d$.  The~EHT has demonstrated that black hole mass measurements for both M87* and Sgr A* can be made by calibrating the scaling relationship between $d$ and $\theta_g$ using a large number of simulated images of the emission structure \citep{M87PaperVI,SgrAPaperIV}.  In~this paper, we thus take $d$ to be a proxy for $M$\footnote{We note that the spin of a black hole also has an effect on the shadow size, but~the impact of spin is small ($\sim$4\%; \citet{Takahashi_2004}) and is dominated by the $\gtrsim$10\% systematic uncertainty associated with the unknown accretion flow morphology \citep{M87PaperVI,SgrAPaperIV}).}; i.e.,~we assume that if $d$ can be measured for a particular SMBH, then $M$ can also be~determined.

\subsection{Proxy for SMBH~Spins}\label{sec:SpinProxy}

The spin, $a$, of~a SMBH has historically proven to be more difficult to measure than the mass; e.g.,~the EHT observations of M87* and Sgr A* have not yet yielded strong constraints on the spin of either SMBH \citep{M87PaperV,SgrAPaperV}.  There are a number of possible avenues for measuring $a$ from horizon-scale images of SMBH systems (e.g., \citep{Ricarte_2022a}), but~the most observationally accessible of these approaches target the imprint of the SMBH spin on the horizon-scale magnetic field topology, which in turn can be accessed through observations of the linear polarization structure around the ring (e.g., \citep{M87PaperVII,M87PaperVIII}).  \citet{Palumbo_2020} have developed a useful decomposition of the polarization structure in terms of a basis that captures the azimuthal behavior of the electric vector position angle (EVPA, i.e.,~the orientation of the linear polarization around the ring).  This decomposition takes the form
\begin{equation}
\beta_m = \frac{1}{S_0} \iint P(r,\phi) e^{-im\phi} r \text{d}r \text{d}\phi ,
\end{equation}
\noindent where $(r,\phi)$ are polar coordinates in the image, $P(r,\phi) = Q(r,\phi) + i U(r,\phi)$ is the complex-valued linear polarization field (with $Q$ and $U$ the standard Stokes intensities), and~$S_0$ is a flux normalization factor.  When studying images of M87* from GRMHD simulations, \citet{Palumbo_2020} found that the ``twisted'' morphology of the linear polarization pattern, quantified by the (complex-valued) $\beta_2$ coefficient, is correlated with the spin of the black hole.  It is now believed that this relation arises from a magnetic field geometry that evolves with the black hole spin: black holes with larger spins exhibit more frame dragging, and~produce more strongly toroidal magnetic fields than lower-spin black holes \citep{Emami_2022}.  \citet{Qiu_2022} further explored the connection between polarized image morphology and SMBH spin, finding that the asymmetry ($A$) of the Stokes $I$ emission, the~polarimetric $\beta_1$ mode, and~the modulus of the polarimetric $\beta_2$ mode also encode spin information (though $\beta_2$ continues to stand out as the most discriminating measurable parameter).  In~this paper, we thus take a joint measurement of $\beta_1$, $\beta_2$, and~$A$ to be our proxy for $a$.

\section{Synthetic Data Generation and Fitting~Procedure}\label{sec:SyntheticData}

To determine the region of the $(\theta, S_{\nu})$ parameter space---and thus the number of SMBHs---for which the quantities of interest described in the previous section could be measured by the ngEHT, we carry out a series of model-fitting exercises using synthetic data.  We use a model for the SMBH emission structure that captures the salient features relevant for measuring the physical quantities of interest.  Per Section~\ref{sec:MeasurableProxies}, these salient features include the diameter and thickness of the emitting ring, as~well as the structure of the linear polarization pattern.  As~our parameterization of the SMBH emission structure, we thus use a polarized ``m-ring'' model \citep{Johnson_2020,SgrAPaperIV} convolved with a circular Gaussian blurring kernel.  This model is restricted to describing ring-like morphologies, but~it can flexibly distribute both the total intensity and the linearly polarized flux about the ring using a relatively small number of parameters.  The~emission structures produced by this model qualitatively match those expected from both simple analytic treatments (e.g., \citep{Gelles_2021}) as well as numerical GRMHD simulations (e.g., \citep{M87PaperVIII,SgrAPaperIV}).

In our polarized source model, the~Stokes I image structure is given by
\begin{equation}
I(r,\phi) = \Bigg[ \frac{S_0}{\pi d} \delta\left( r - \frac{d}{2} \right) \sum_{k=-m}^m \alpha_k e^{i k \phi} \Bigg] * \Bigg[ \frac{4 \ln(2)}{\pi W^2} \exp\left( -\frac{4 \ln(2) r^2}{W^2} \right) \Bigg] ,
\end{equation}
\noindent where $*$ denotes the convolution operation, $d$ is the ring diameter, $W$ is the FWHM ring width, and~$\delta$ denotes the Dirac delta function.  We enforce $\alpha_0 = 1$ so that $S_0$ is the total flux density, and~we also enforce $\alpha_{-k} = \alpha_k^*$ so that the image intensity is real-valued.  We define $A = |\alpha_1|$ to be the asymmetry parameter mentioned in Section~\ref{sec:MeasurableProxies} as potentially relevant for spin constraints.\footnote{Note that this definition for $A$ differs from that in \citet{Qiu_2022}, who instead adopt the asymmetry definition used in \citet{Medeiros_2022}.}  The linear polarization structure is similarly given by
\begin{equation}
P(r,\phi) = \Bigg[ \frac{1}{\pi d} \delta\left( r - \frac{d}{2} \right) \sum_{k=-m}^m \beta_k e^{i k \phi} \Bigg] * \Bigg[ \frac{4 \ln(2)}{\pi W^2} \exp\left( -\frac{4 \ln(2) r^2}{W^2} \right) \Bigg] ,
\end{equation}
\noindent where we now allow both $\beta_{-k}$ and $\beta_{k}$ to be free parameters because $P$ is complex-valued in~general.

\begin{figure}[H]
\centering
\includegraphics[width=0.6\columnwidth]{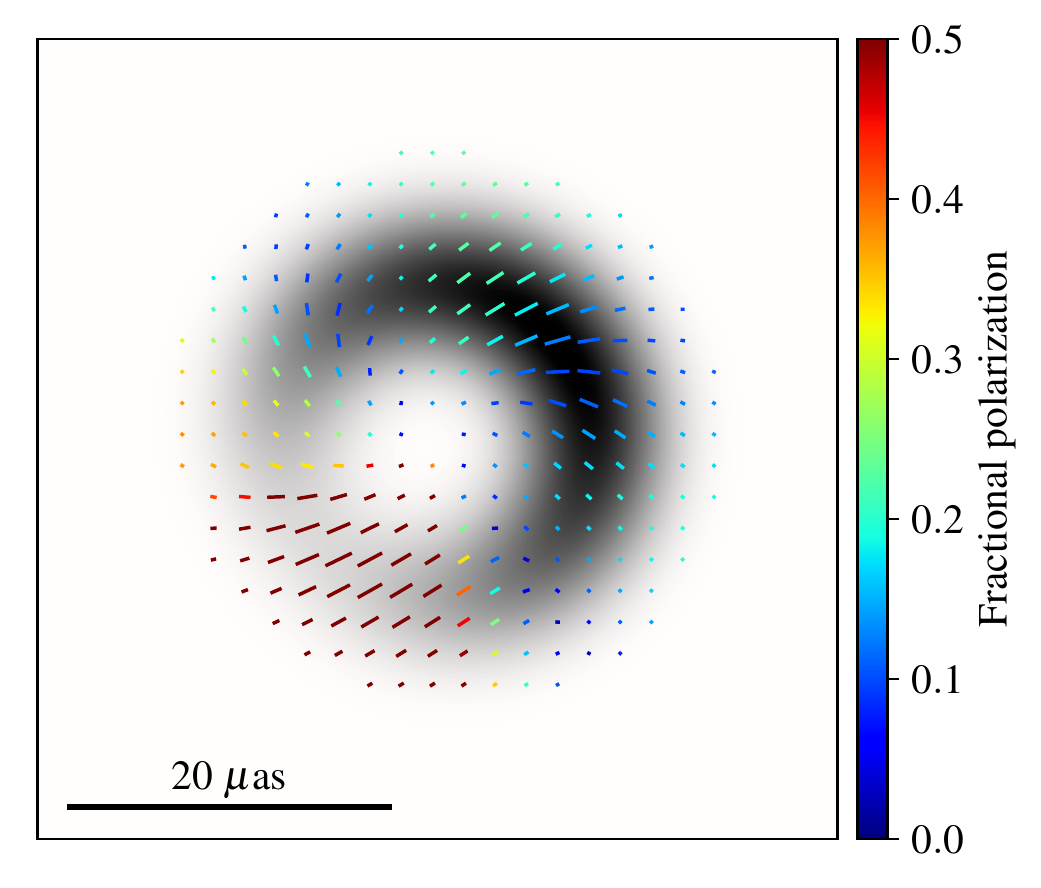}
\caption{Example polarized source model used for generating the synthetic data described in Section~\ref{sec:SyntheticData}.  The~grayscale image shows the Stokes I emission, while the colored ticks mark the EVPA of the linear polarization structure.  The~length of each tick is proportional to the intensity of the linear polarization (i.e., $|P|$), while the color of each tick reflects the fractional polarization (i.e., $|P|/I$).} \label{fig:model}
\end{figure}

We generate a number of synthetic SMBH images by gridding the $($d$, S_{0})$ parameter space, spanning $[0.1,100]$\,$\mu$as in $d$ and $[10^{-3},1]$\,Jy in $S_0$, with both dimensions uniformly gridded on a logarithmic scale.  We set $m=1$ for the Stokes I emission, with~both the real and imaginary parts of $\alpha_1$ uniformly sampled within $[-0.5,0.5]$.  For~the polarized emission we set $m=2$, with~the real and imaginary parts of $\beta_0$ and $\beta_{-2}$ uniformly sampled within $[-0.1,0.1]$, the~real and imaginary parts of $\beta_1$ and $\beta_{-1}$ uniformly sampled within $[-0.05,0.05]$, and~the real and imaginary parts of $\beta_2$ uniformly sampled within $[-0.3,0.3]$.  For~all synthetic images, we enforce $W = d/3$.  Though~these choices are not unique, they cover a range of parameter values similar to that seen in the GRMHD simulations developed by the EHT collaboration \citep{M87PaperV,SgrAPaperV}.  An~example polarized m-ring image generated using these specifications is shown in \autoref{fig:model}.

To generate synthetic ngEHT observations corresponding to the synthetic images, we use the \texttt{ngehtsim}\footnote{\url{https://github.com/Smithsonian/ngehtsim}} package, which expands on the synthetic data generating functionality of the \texttt{ehtim} library \citep{Chael_2016,Chael_2018}.  We assume the observations are carried out at an observing frequency of 230\,GHz and with 8\,GHz of bandwidth using the ``full'' ngEHT Phase 1 array configuration from \citep{RefArray}, which consists of the 2022 EHT array plus the OVRO 10.4\,m dish, the~Haystack 37\,m dish, and~three 6.1\,m dishes located in Baja California (Mexico), Las Campanas Observatory (Chile), and~the Canary Islands (Spain).  We use historical weather data to determine appropriate system equivalent flux densities at each site following a procedure similar to that in \citet{Raymond_2021}.  To~emulate fringe-finding signal-to-noise ratio (SNR) thresholds, we flag any visibilities from baselines that contain a station not participating in at least one other baseline that achieves an SNR of 5 in a 10-s integration time.  We add complex station gain corruptions at the level of 10\% in amplitude and uniformly sampled within $[0,2\pi]$ in phase for all stations on every 300-second time interval, to~emulate scans, and~we assume that the data have been calibrated to remove polarimetric leakage~effects.

We generate synthetic datasets across a grid in right ascension and declination, with~spacings between grid points of 1 hour in right ascension and 10 degrees in declination.  To~gather information on the performance of the array in different weather conditions and for different black hole structure realizations, we generate 100 instantiations of synthetic data at each grid location.  We assume weather conditions typical for the month of~April.

For each synthetic dataset, we estimate the precision with which the parameters of a polarized m-ring model fit to the data could be recovered.  We compute these estimates using a Fisher matrix approach implemented within the \texttt{ngEHTforecast}\footnote{\url{https://github.com/aeb/ngEHTforecast}} package.  This approach does not explicitly carry out fits of the model to the data; instead, it assumes that a ``good'' fit to the data has already been achieved, and~it then provides an estimate of the uncertainty in each of the fitted parameters via a second-order expansion of the logarithmic probability density around the best-fit location.  We compute parameter precision estimates assuming that the fits have been carried out using complex visibilities as the input data products, with~broad priors on the station gain amplitudes and phases at every~scan.

\section{Results: The Expected Number of Measurable SMBH Masses, Spins, and~Shadows}\label{sec:NumberOfMeasurables}

\begin{figure}[t]
\centering
\includegraphics[width=0.8\columnwidth]{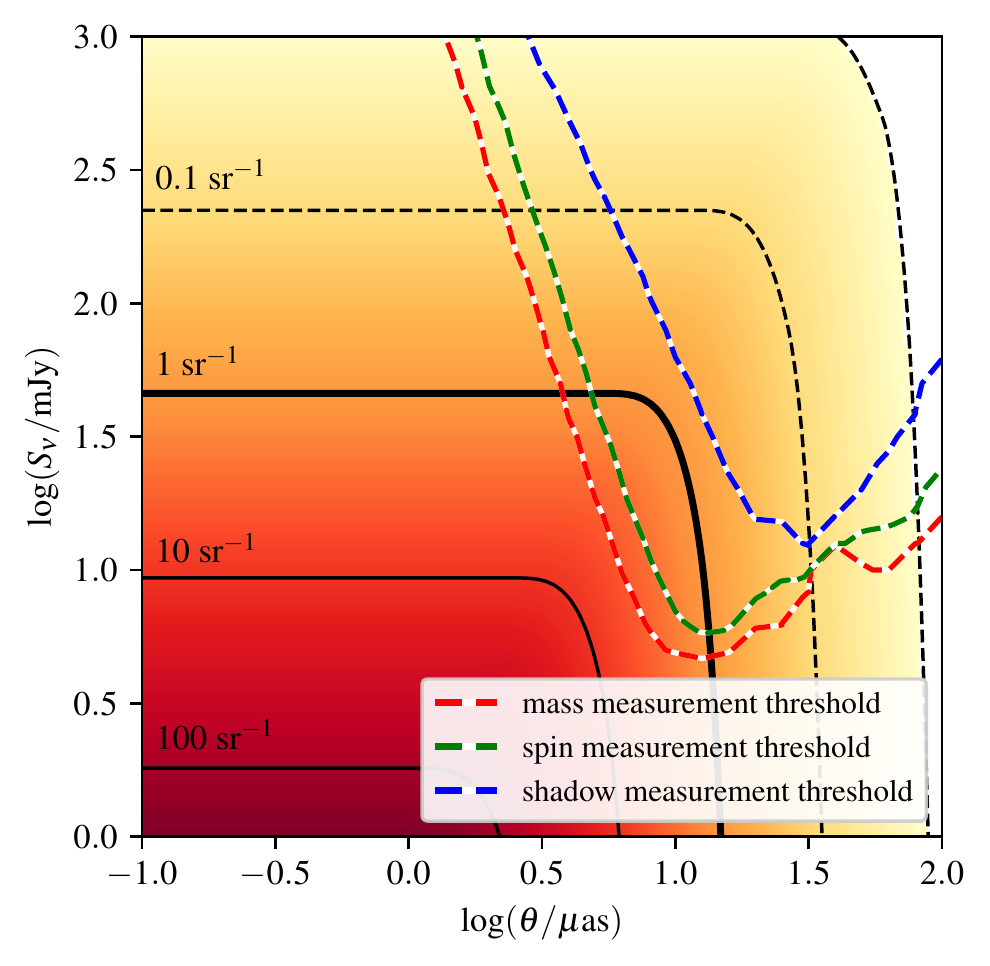}
\caption{Approximate number density of SMBHs that are expected to satisfy different thresholds of measurability, assuming an observing frequency of 230\,GHz.  The~background colorscale and contours mark the number density (per unit solid angle) of SMBHs that have flux densities greater than $S_{\nu}$ and shadow diameters larger than $\theta$, as~a function of $S_{\nu}$ and $\theta$ and assuming that sources are distributed isotropically on the sky \citep{Pesce_2021}.  The~solid contours start with the thick contour indicating a count of 1 and then increase by factors of 10 towards the lower left, while the dashed contours each decrease by a factor of ten towards the upper right.  The~overplotted colored dashed contours indicate where various parameters of interest could be measurable for different combinations of $(\theta,S_{\nu})$, assuming observations appropriate for the ``full'' ngEHT Phase 1 array observing at a declination of 10 degrees (i.e., averaged over right ascension).  The~red dashed contour marks the lower boundary of the region in which black hole mass can be measured, the~green dashed contour marks the lower boundary of the region in which black hole spin can be measured, and~the blue dashed contour marks the lower boundary of the region in which black hole shadow can be~measured.} \label{fig:measurement_thresholds}
\end{figure}

\begin{figure}[p]
\centering
\includegraphics[width=1.0\columnwidth]{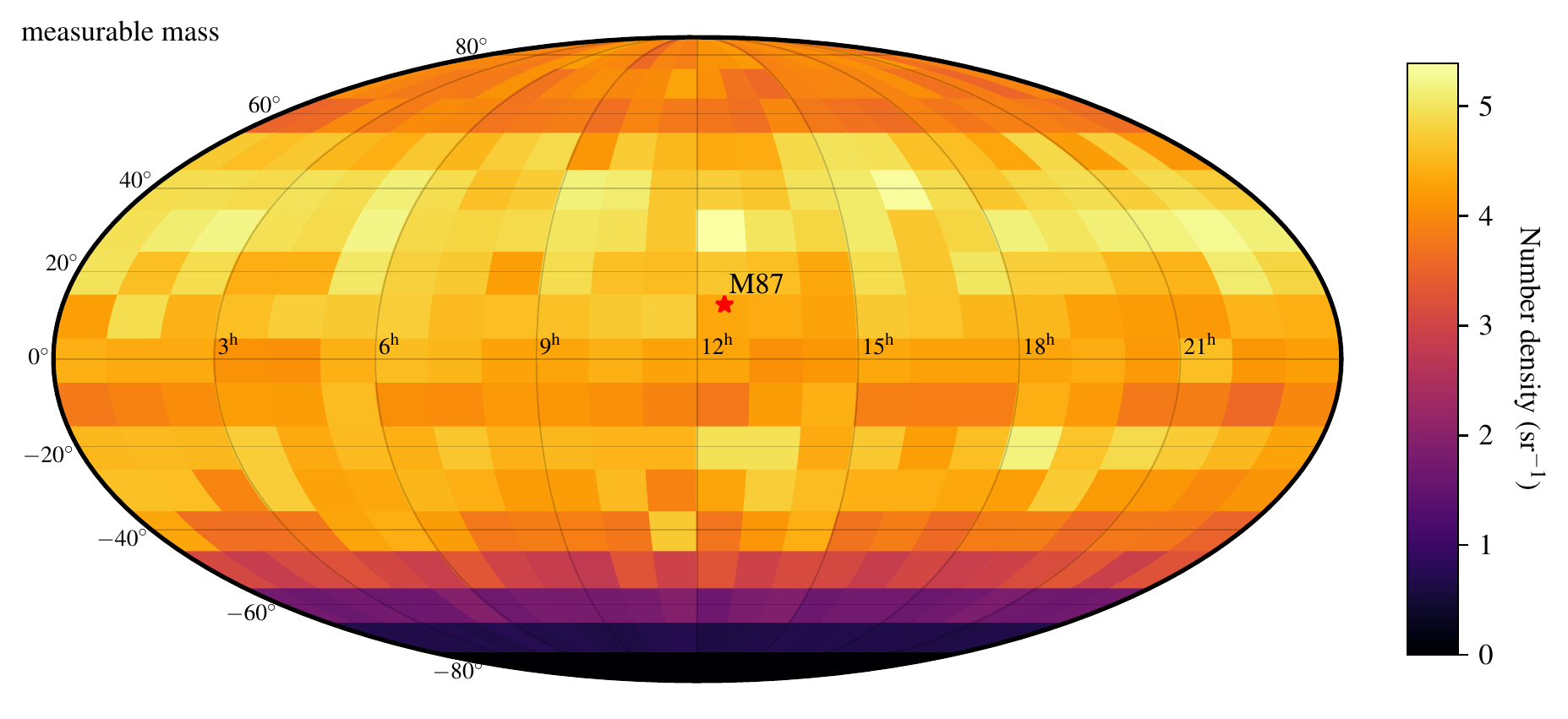}
\includegraphics[width=1.0\columnwidth]{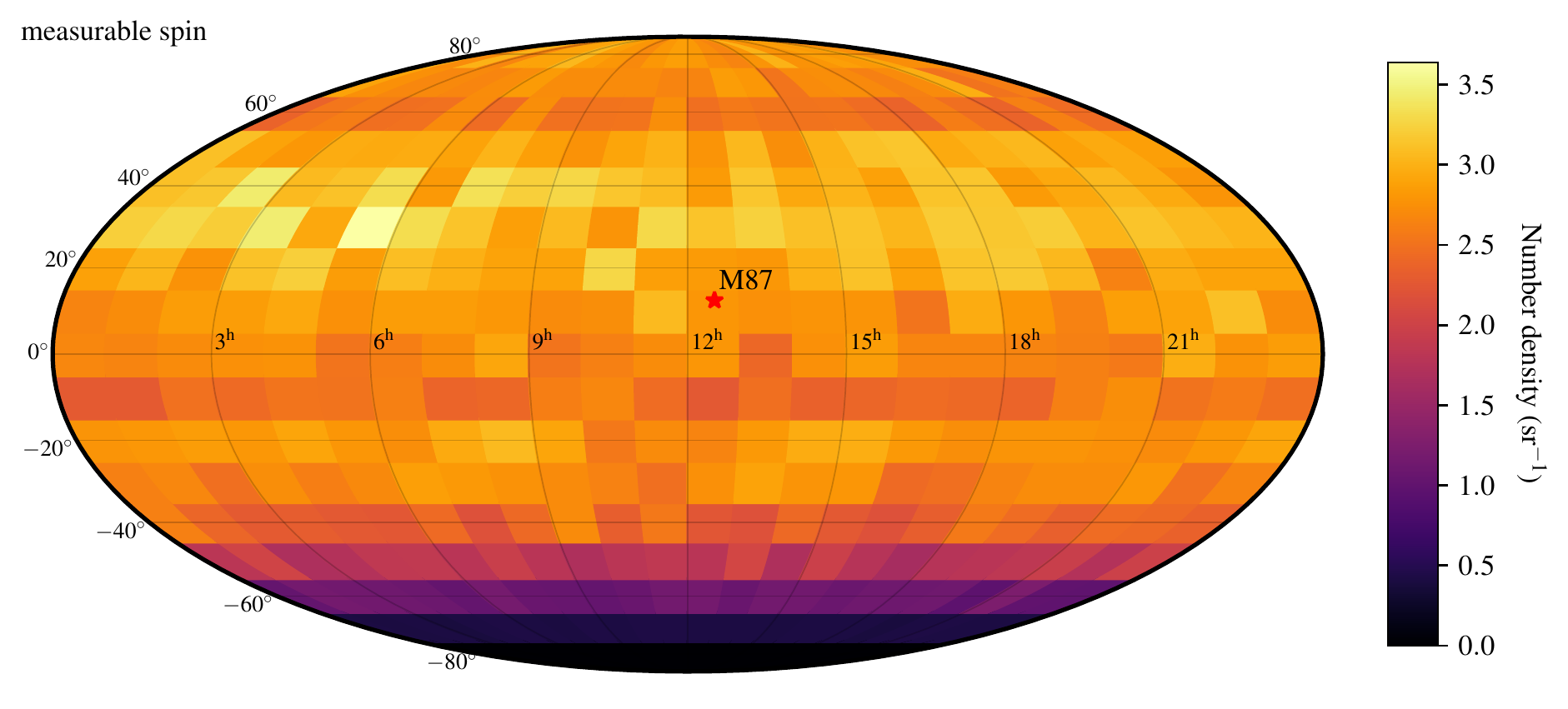}
\includegraphics[width=1.0\columnwidth]{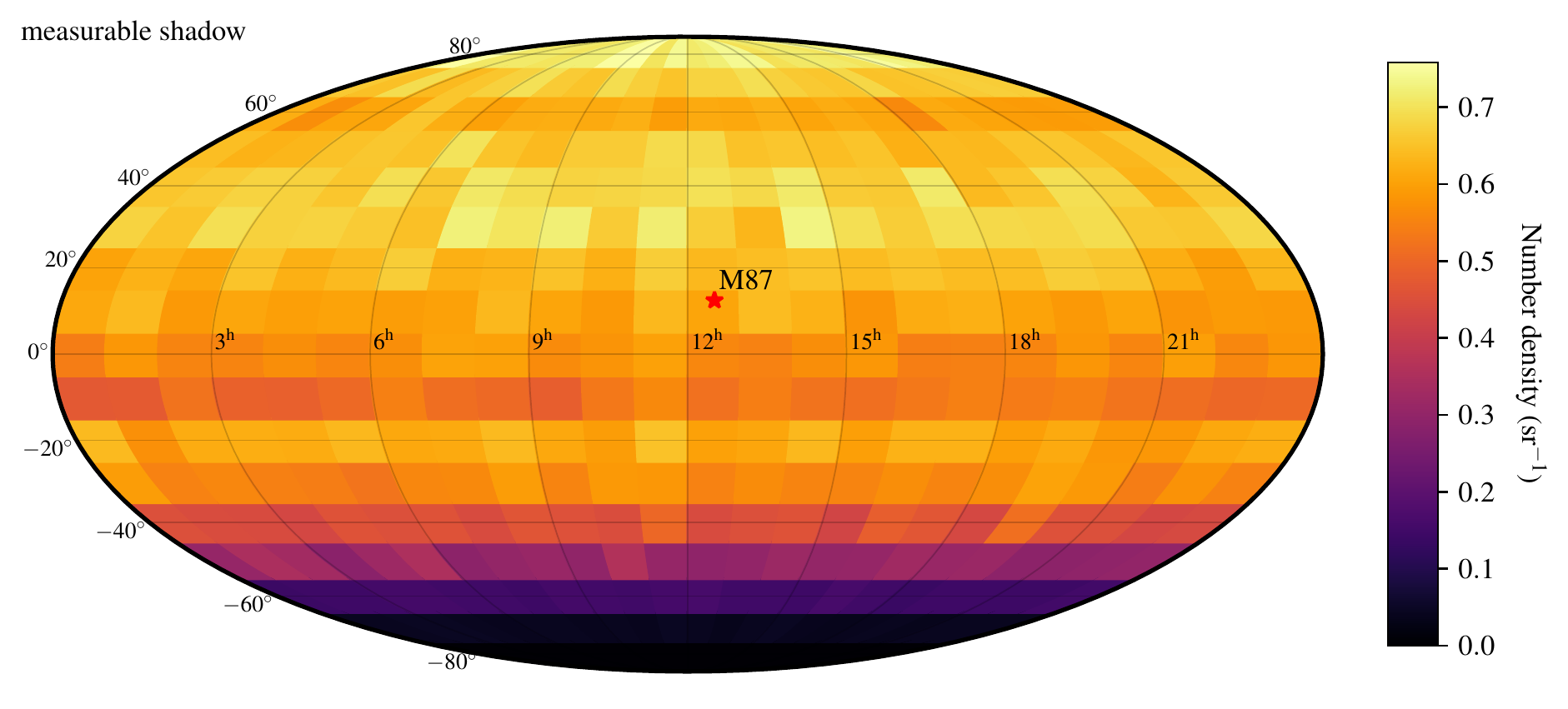}
\caption{Estimated sky density of SMBHs with measurable masses (top), spins (middle), and~shadows (bottom), as~a function of right ascension and declination.  These estimates have been determined according to the criteria outlined in Section~\ref{sec:NumberOfMeasurables}, and~they assume an underlying distribution of observable SMBHs from \citet{Pesce_2021}.  The~stochastic variations seen from pixel to pixel are primarily the result of sampling noise.  The location of M87* is marked with a red~star.} \label{fig:SourceCounts}
\end{figure}

The results of the modeling exercises described in the previous section are summarized in \autoref{fig:SourceCounts}, which shows the sky density of sources expected to have measurable masses (top panel), spins (middle panel), and~shadows (bottom panel).  At~each sky location, the~plotted density corresponds to an average over 100 instantiations of weather conditions and source structure.  Our criteria for determining whether a particular mass, spin, or~shadow is deemed ``measurable'' are as follows:

\begin{enumerate}
    \item Our condition for whether a SMBH has a measurable mass is that the fractional uncertainty in the measurement of the ring diameter $d$ must be at the level of 20\% or lower (i.e., it is measured with a statistical significance ${\gtrsim} 5 \sigma$).  Values of $(\theta,S_{\nu})$ for which this condition is satisfied fall to the upper right of the red dashed curve in \autoref{fig:measurement_thresholds}.
    \item Our condition for whether a SMBH has a measurable spin is that the uncertainty in the measurement of all spin-relevant parameters (as determined by \citet{Qiu_2022}; see also Section~\ref{sec:SpinProxy}) must be at the level of 20\% or lower.  Specifically, we require the fractional uncertainty in $|\alpha_1|$, $|\beta_1|$, and~$|\beta_2|$ and the uncertainty in $\text{arg}(\beta_1)$ and $\text{arg}(\beta_2)$ to all be less than 0.2 (i.e., 20\%).
    Values of $(\theta,S_{\nu})$ for which this condition is satisfied fall to the upper right of the green dashed curve in \autoref{fig:measurement_thresholds}.
    \item Our condition for whether a SMBH has a measurable shadow is that the fractional width $W/d$ deviates from unity with an uncertainty of 20\% or smaller; i.e.,~we require that $W < d$ with a statistical significance ${\gtrsim} 5 \sigma$.
    Values of $(\theta,S_{\nu})$ for which this condition is satisfied fall to the upper right of the blue dashed curve in \autoref{fig:measurement_thresholds}.
\end{enumerate}

Given the above measurability thresholds, we can see from \autoref{fig:measurement_thresholds} that there is a hierarchy of measurement difficulty with increasing $S_{\nu}$ and $\theta$.  The~``easiest'' quantity to measure is $d$ (and thus the black hole mass), which can be recovered for $\sim$50 sources after integrating over the whole sky.  The~next most well-constrained quantities are those pertaining to the black hole spin, which we find can be recovered for $\sim$30 sources.  The~most difficult quantity to measure is $W$ (and thus the black hole shadow), which can be recovered for $\sim$7 sources.  The~measurements are cumulative within this hierarchy: for all sources for which spin is measurable, mass is also measurable; for all sources for which the shadow is measurable, both spin and mass are also~measurable.

All three quantities of interest exhibit two regimes of non-measurability in \autoref{fig:measurement_thresholds}.  For~sources with flux densities below $S_{\nu} \lesssim 10$\,mJy, the~source is too weak to be detected on most baselines, and~there are thus simply insufficient data to enable significant constraints on the parameters of interest.  For~sources that are stronger than $\sim$10\,mJy but smaller than several $\mu$as, there can be many detected data points, but~the source is insufficiently resolved to enable significant constraints on morphological parameters.  In~both regimes, all three quantities of interest exhibit a measurability tradeoff between $\theta$ and $S_{\nu}$.  In~the second regime, this tradeoff is such that it is possible to make a measurement for sources with smaller $\theta$ so long as they have sufficiently larger $S_{\nu}$ (because increasing signal-to-noise ratio permits subtler features to be recovered), while in the first regime the tradeoff is reversed (because compact sources yield more detections---particularly on long baselines---than extended sources).

\autoref{fig:SourceCounts} shows the sky distribution of objects with measurable masses, spins, and~shadows, after~averaging over weather and source structure instantiations.  We find that the distribution is quite uniform, and~that accessible objects can be located almost anywhere in the sky; there is no strong dependence on right ascension.  The~only major gaps in accessibility are for sources having declinations within $\sim$30 degrees of the southern celestial pole, for~which the $(u,v)$-coverage of the array is particularly poor.  A~modest increase in source density is seen around declinations of $\sim$30--40 degrees, where the $(u,v)$-coverage of the array is densest.  For~the shadow measurements, we also see a modest increase in source density around the northern celestial pole; northern polar observations provide the most complete long-baseline coverage, so this bump in density may indicate that the long baselines are the most constraining for the width~parameter.

\section{Summary and~Conclusions}\label{sec:Summary}

To date, the~EHT has observed the horizon-scale emission structure around two SMBHs.  The~ngEHT aims to improve on the capabilities of the EHT by adding new dishes to the array, increasing the observing bandwidth, and~expanding the frequency coverage, all of which will improve the sensitivity and fidelity of reconstructed images.

Motivated by the promise of the ngEHT for population studies of SMBHs, we have identified three scientific quantities of interest that the ngEHT can expect to measure for a number of SMBHs: the black hole mass, the~black hole spin, and~the black hole shadow.  We construct a geometric ring model for the polarized emission structure around a SMBH, and~we identify parameters of this model as observable proxies for the scientific quantities of interest.  Specifically, we associate the diameter of the ring with measurements of the black hole mass, the~thickness of the ring with measurements of the black hole shadow, and~the linear polarization structure with measurements of the black hole~spin.

Assuming a Phase 1 ngEHT array configuration observing in April conditions at a frequency of 230\,GHz, we generate a large number of realistic synthetic observations spanning a range of source structure (i.e., flux density $S_{\nu}$ and angular size $\theta$) and site weather (i.e., opacity and atmospheric temperature) instantiations.  For~each synthetic dataset, we use a Fisher matrix formalism to estimate the precision with which each of the geometric ring model parameters of interest could be measured.  We use the statistics of these measurement precision estimates (across all weather instantiations) to determine the corresponding number of SMBHs on the sky whose properties could be well-constrained as a function of $S_{\nu}$ and $\theta$.  We carry out this procedure for synthetic observations covering a grid in right ascension and declination, finding that the sky density of measurable sources (in each parameter of interest) is approximately uniform for declinations above roughly $-60^{\circ}$.

Associating these measurable parameters with their corresponding physical quantities of interest, we present estimates for the number of SMBHs for which the Phase 1 ngEHT can expect to make measurements of these quantities.  Integrating over the whole sky, we find that the Phase 1 ngEHT should be able to measure $\sim$50 black hole masses, $\sim$30 black hole spins, and~$\sim$7 black hole shadows.  The~measurable SMBHs have characteristic observed flux densities of $\sim$30\,mJy and angular sizes of $\sim$10\,$\mu$as; per \citet{Pesce_2021}, we expect the bulk of these SMBHs to lie in the redshift range between $z \approx 0.01$ and $z \approx 0.1$.  Our estimate for the number of measurable shadows is consistent with the predictions from \mbox{\citet{Pesce_2021}}.

We note that our detection criteria for mass and spin are likely optimistic. A~primary analysis limitation is that our model for the appearance of an SMBH does not include emission that extends much beyond the near-horizon region.  Mass estimates for SMBHs of interest to the ngEHT may be complicated by additional image features---such as, e.g.,~AGN jets (e.g., \citep{Janssen_2021}---that could limit the ability to accurately estimate the ring diameter when it is only marginally resolved or weakly detected.  For~spin, the~situation is even more uncertain: the EHT has already produced tight estimates for the ring parameters $\beta_1$, $\beta_2$, and~$A$ for M87*, but~it has not yet claimed a corresponding measurement of the black hole spin \citep{M87PaperVI,M87PaperVII,M87PaperVIII}.  A~secure association between these ring parameters and spin will require a combination of continued observational and theoretical~studies.

On the other hand, we have employed a simplified analysis that likely underestimates the number of accessible sources, given any particular set of detection criteria.  For~instance, the~synthetic datasets used in this paper are currently limited to April weather conditions and an observing frequency of 230\,GHz; a more comprehensive exploration of year-round weather conditions and the addition of a 345\,GHz observing band would likely increase the number of accessible sources.  Furthermore, the~synthetic datasets generated for the analyses in this paper have assumed an EHT-like calibration procedure; more advanced calibration strategies that can bootstrap phase information across frequency bands (e.g.,~\citep{Rioja_2020}) are also expected to increase the number of accessible sources.  Addressing these shortcomings will be the focus of future~work.

Observationally, the~most critical next step is to identify a list of credible targets and start surveying them to determine flux densities and compactness for ngEHT followup.  \citet{Ramakrishnan_2022} are compiling a comprehensive sample of all plausible ngEHT AGN targets, which is expected serve as a source catalog for pursuing SMBH population studies with the~ngEHT.

\vspace{6pt} 

\authorcontributions{Conceptualization, J.L.G., P.N., D.W.P.; methodology, M.D.J., D.C.M.P., D.W.P., A.R.; software, A.E.B., M.D.J., D.W.P.; writing---original draft preparation, D.W.P.; writing---review and editing, A.E.B., J.L.G., M.D.J., N.M.N., P.N., D.C.M.P., D.W.P., A.R.; visualization, D.W.P.; supervision, J.L.G., P.N.  All authors have read and agreed to the published version of the~manuscript.}

\funding{Support for this work was provided by the NSF through grants AST-1440254, AST-1935980, and~AST-2034306, and~by the Gordon and Betty Moore Foundation through grant GBMF-10423.  This work has been supported in part by the Black Hole Initiative at Harvard University, which is funded by grants from the John Templeton Foundation and the Gordon and Betty Moore Foundation to Harvard University. NN acknowledges funding from TITANs NCN19-058 and Fondecyt~1221421.}

\conflictsofinterest{The authors declare no conflict of~interest.}

\begin{adjustwidth}{-\extralength}{0cm}
	\printendnotes[custom]

\reftitle{References}

\end{adjustwidth}
\end{document}